\documentclass[11pt]{article}
\newcommand{\be}{\begin{equation}}
\newcommand{\ee}{\end{equation}}
\newcommand{\beq}{\begin{eqnarray}}
\newcommand{\eeq}{\end{eqnarray}}

\usepackage{graphicx}
\begin{document}
\begin{center}
{\bf\LARGE QED Vacuum Loops and Dark Energy }\\
[7mm]
\vspace{1cm}  
H. M. FRIED
\\
{\em Department of Physics \\
Brown University \\
Providence R.I. 02912 USA}\\
fried@het.brown.edu\\
[5mm]
Y. GABELLINI
\\
{\em Institut Non Lin\'eaire de Nice\\
UMR 7335 CNRS\\
 1361 Route des Lucioles\\
06560 Valbonne France}\\
yves.gabellini@unice.fr\\
[5mm]

\vspace{4cm}
Abstract
\end{center}
A QED--based `` bootstrap '' mechanism is suggested as an explanation for the vacuum energy that furnished the initial impulse for Inflation, and continues on to provide present day  Dark Energy. Virtual vacuum fluctuations are assumed to generate effective electromagnetic fields whose average value corresponds to an effective c--number $A_{\mu}^{\rm vac}(x)$, which is itself equal to the vacuum expectation value of the operator $A_{\mu}(x)$ in the presence of that  $A_{\mu}^{\rm vac}(x)$. Lorentz invariance is manifest, as every observer would measure the same electric field in his or her own reference frame. The model has one arbitrary parameter $\xi$, and fits the energy density of present day Dark Energy for $\xi\sim O(1)$.
\newpage
{\bf\section{Introduction}}
\setcounter{equation}{0}
Every so often, one is struck by an idea which is so simple and compelling that it is hard to understand why it has been overlooked. The essence of this very simple idea is as follows. Conventionally, electromagnetic fields are either `` quantized '' or `` external '', and by external is meant classical fields which can be switched on and off. Simultaneously, one speaks of the `` quantum vacuum~'', in which quantized fields of arbitrary complexity are fluctuating, and whose effects can only be indirectly inferred ( such as the 27 Megacycles of the famous 1057 Megacycle Lamb shift ).

Imagine that one has available a `` super Heisenberg '' gamma--ray microscope, so that one could `` see '' a virtual photon of the vacuum suddenly transform itself into a bubble  corresponding to the virtual appearance of an electron and positron, whose propagations define the sides of the bubble, and which propagation continues for a duration on the order of the inverse of their mass. When these virtual particles are separated there is an electric field between them, which can be thought of an electromagnetic fluctuation that disappears when the bubble collapses. Such fluctuations are present in the vacuum; they cannot be turned off and, in an appropriately averaged way, they should contain energy. They arise from the fundamental quantization of the operator QED fields. The averaged or background electromagnetic energy of these fluctuations might well be characterized by the existence of an effective, c--number field $A_{\mu}^{\rm vac}$ which is always present. The existence of such an averaged, effective field, due to its quantum origin, could possibly have striking effects on the classical world about us; but because of the form and high frequency of such an $A^{\rm vac}(x)$, it should have no observable effect on the motion of classical charged particles.

 In this paper, we shall explain, in some detail, the initial mechanism which could account for a simple understanding of Dark Energy, corresponding to a new,   `` bootstrap '' solution in QED. A previous arXiv submission \cite{one} described a different solution; the present, ``~improved~'' solution provides a simple, physical picture for the present day Dark Energy.

One prejudice of the authors should be stated at the outset, which forms part of the motivation for the present remarks. A conventional approach to vacuum energy is that the latter represents, in some fashion, zero point energies of relevant quantum fields. Aside from being divergent, and so requiring renormalizations, it is not clear that zero point terms even belong in any field theory, for they can be well understood as the remnants of improper positioning of products of operators at the same space--time point. One cannot proceed from classical to quantum forms without facing this question, which has long ago been given a Lorentz invariant answer by Symanzik \cite{two} in terms of `` normal ordering ''. When Lorentz invariance is subsumed into a more general relativistic invariance, in which all energies couple to the metric, there is still no requirement of including those zero point terms which should have been excluded at the very beginning; instead, zero point terms have been pressed into service as the simplest way of attempting to understand vacuum energy.

Because zero point energies have been used to give a qualitative explanation of the logarithm of the lowest order Lamb shift does not mean that they are the correct explanation for that effect; zero point energies cannot produce the additive constant to that logarithm, and they most certainly cannot produce the higher order terms which have been experimentally verified. It is here suggested that there is another source of vacuum energy, one which violates no principles of quantum field theory, that could provide a simple and qualitatively reasonable QED mechanism for Dark Energy.

At this point, it may be useful to consider just what is meant by a Vacuum Energy, especially in order to understand what is meant by `` effective Lorentz invariance ''. In the context of our Model, one imagines charged fermion loops fluctuating in and out of existence, with the normal vector to the plane of their loop fluctuating in all possible directions. At the instant that any given loop appears, there is an electric field across that loop, between the oppositely charged fermions forming the sides of the loop; and that field signifies that at that instant there is a certain amount of electrostatic energy present, which apparently disappears when the loop collapses. But then another loop, with its normal vector pointing in a different direction appears; and then another, and another, such that it is clear that, on average, there is a certain amount of energy associated with such continuously random fluctuations.

We phrase the question in terms of an effective c-number field,  $A_{\mu}^{\rm vac}(x)$, whose polarization vector can only lie in the time direction, $\mu = 0$, because of the random orientations of the fluctuating planes, as imagined by every observer in his own Lorentz frame. And we find that each such observer imagines the same form and magnitude of this vacuum energy. Note that the word `` imagines '' is used, rather than the more usual `` sees '' or `` measures~'', and this is because such an effectively invariant vacuum energy cannot be described by conventional Lorentz transformations that are used to describe classical fields and particle transformations. This is an energy, an effective potential energy, that one cannot measure with ordinary equipment, especially since its frequency -- the $M$ value -- is so very large, see eq.(5.3). But there $\underline {\rm is}$ an energy present, an average potential energy of all the fluctuating loops, which can have significant effects on a large, classical scale. This point will be discussed in more details in Section 4.

As noted above, we propose this new form of effective potential energy as a real effect, due to the quantum fluctuations of the quantized fields; and we offer it as a possible example of the source of the present day acceleration of the Universe. Because the shape of this effective potential energy at small distances ( or times ) is so suggestive of Inflation, whose treatment involves a significant variation from that provided in the present paper, but whose essence is of the same origin, the subject of Inflation will be treated in a subsequent publication.

Finally, we note that this effective potential energy corresponds to the result of loop fluctuations with the normal vector to the plane of the loop continuously and randomly appearing in arbitrary spatial directions, and with a very high frequency given by the derived parameter $M$. It then becomes clear why such an effective potential energy could never initiate a Schwinger mechanism \cite{four}, which would tear the fermionic loop apart, converting their constituents into real, propagating particles : that mechanism requires a reasonably constant ( and strong ) electric field, rather than a rapidly oscillating one.
\bigskip
{\bf\section{Formulation}}
\setcounter{equation}{0}
We take the point of view that such an averaged $A_{\mu}^{\rm vac}(x)$ can be treated as an effective, classical, external field, albeit one which is always present, and should be included in QED considerations, and especially in the construction of the QED Generating Functional (GF). There, the Schwinger--Symanzik--Fradkin formulations \cite{three} produce a GF, ${\cal Z}[\eta,\bar\eta, J]$ : 
$${\cal Z}[\eta, \bar\eta, J] = \,<\!0\,|\Bigl(e\,^{\displaystyle\!i\int\!\!(\bar\eta\psi+\bar\psi\eta + JA)}\Bigr)_+|\,0\!>$$
 which, for our purposes, can be transformed into the most convenient form :
\beq\matrix{\displaystyle&{\cal Z}[\eta, \bar\eta,J] =\displaystyle {\cal N}\,\exp\Bigl[ { i\over2}\int\!\!d^4x\,d^4y\,J^{\mu}(x)\,D_{c,\mu\nu}(x-y)\, J^{\nu}(y)\Bigr]\hfill\cr\noalign{\medskip} & \displaystyle\times e\,^{\displaystyle {\cal D}\!_A}\exp\Bigl[i\!\int\!\!d^4x\,d^4y\,\bar\eta(x)\, G_c(x,y|A)\,\eta(y)\Bigr]\exp\bigl[ L[A]\bigr]\,\hfill}\eeq
with $\displaystyle A_{\mu}(x) = \int\!\!d^4y\,D_{c,\mu\nu}(x-y)\,J^{\nu}(y)$. 

The relativistic notation used throughout this article is the so--called east coast metric, with the scalar product defined by $a\!\cdot\!b = \vec a\cdot\vec b - a_0b_0$, and the Dirac matrices such that $\gamma_{\mu}^{\dagger} = \gamma_{\mu}$, $\gamma_{\mu}^2 = 1$ and $\{\,\gamma_{\mu}, \gamma_{\nu}\,\} = 2\delta_{\mu\nu}$.

The normalization ${\cal Z}[\,0, 0, 0\,] = 1$ is defined by : 
${\cal N}^{-1} = <\!0\,|S|\,0\!>\,=\,e\,^{\displaystyle {\cal D}\!_A}\,\exp\bigl[ L[A]\bigr]\Bigl|_{A=0}$, $G_c[A] = \bigl[ m +\gamma\!\cdot\!(\partial - ieA)\bigr]^{-1}$ is the Feynman ( causal ) electron propagator in an arbitrary external field $A_{\mu}(x)$; $ L[A] = {\rm Tr}\ln\bigl(\, G_c^{-1}[A]G_c[0]\,\bigr)$ is the vacuum functional corresponding to a single closed lepton loop, to which is attached all possible ( even ) number of fields $A_{\mu}$; and $D_{c}(x-y)$ is the free photon propagator in an arbitrary relativistic gauge. We shall refer to the quantity :
$$e\,^{\displaystyle {\cal D}\!_A}\ \ \ {\rm with}\ \ \ \ \displaystyle{\cal D}\!_A = -{i\over2}\,\int\!\!d^4x\,d^4y\,{\delta\over\delta A_{\mu}(x)}D_{c,\mu\nu}(x-y){\delta\over\delta A_{\nu}(y)}$$
as the linkage operator, for its function is to link all pairs of $A$--dependence upon which it acts by virtual $D_c$ propagators. We begin this discussion specifying those radiative corrections due to a virtual electron--positron bubble, but will later generalize to the corresponding situations when similar contributions are made by virtual bubbles of other particles.

Eq. $(2.1)$ as written is true when no classical, external field is present; however if such fields were present, $(2.1)$ would still be true if the $A_{\mu}$ in $G_c$ and $L$ were replaced by $A_{\mu} + A_{\mu}^{\rm ext}$. In this way, it is easy to see the relation between QED  and ordinary Quantum Mechanics : were the $A$ dependence and the linkage operator suppressed, one has the GF for many--body Potential Theory, where the fermions are moving in the potential $A_{\mu}^{\rm ext}$. All of the virtual structure of QED corresponds to the action of that linkage operator, connecting all the $A$ dependence upon which it acts, in all possible ways.

Let us now assume the existence of an effective, classical field arising from the virtual fluctuations of the QED vacuum; for the present discussion, no ordinary, classical, external field need be included. Conventionally, the vacuum expectation value ( vev ) of the current operator $j_{\mu}(x) = ie\bar\psi(x)\gamma_{\mu}\psi(x)$ must vanish in the absence of classical, external fields, designated by $A_{\mu}^{\rm ext}(x)$ :
$$<\!0\,|\,j_{\mu}(x)|\,0\!>\!\Bigl|_{A^{\rm ext}=0} =0$$
When such a classical $A_{\mu}^{\rm ext}$ is present, the current it induces in the vacuum can be non zero~:
$$<\!0\,|\,j_{\mu}(x)|\,0\!>\!\Bigl|_{A^{\rm ext}\neq 0} \neq 0$$
and strict current conservation demands that $\partial_{\mu}\!<\!0\,|\,j_{\mu}(x)|\,0\!>\!\Bigl|_{A^{\rm ext}\neq 0} = 0$.

The conventional mathematical apparatus used to describe the vacuum state, or vev of currents operators, makes no reference to the scales on which vacuum properties are to be observed; and in this sense, it is here suggested that the conventional description is incomplete. On distance scales larger than the electron's Compton wave lenght, $\lambda_e\sim10^{-10}$ cm, one can imagine that the average separation of virtual $e^+$ and $e^-$ currents is not distinguishable, and hence that the vacuum displays not only a zero net charge, but also a zero charge density. But on much smaller scales, such as $10^{-20}$ cm, the average separation distances between virtual $e^+$ and $e^-$ are relatively large, with such currents describable as moving in each other's field -- until they annihilate. Since there is nothing virtual, or off shell, about charge, on sufficiently small space--time scales such `` separated '' currents can be imagined to produce `` effective c--number '' fields, characterized by an $A_{\mu}^{\rm vac}(x)$, which could not be expected to be measured at distances larger than $\lambda_e$, but which exist and contain electromagnetic energy on scales much smaller than $\lambda_e$. We therefore postulate that, at such small scales and in the absence of conventional, large scale $A_{\mu}^{\rm ext}$, $<\!0\,|\,j_{\mu}(x)|\,0\!>$ need be neither $x$--independent nor zero; but rather, that it generates an $A_{\mu}^{\rm vac}(x)$ discernible at such small scales, which is given by the conventional expression :
\beq A_{\mu}^{\rm vac}(x) = \int\!\!d^4y\,D_{c,\mu\nu}(x-y)\,<\!0\,|\,j_{\nu}(y)|\,0\!>\eeq
where $D_{c,\mu\nu}$ is the usual, free field, Feynman photon propagator that, for convenience, is defined in the Lorentz gauge, $0 = \partial_{\mu}A_{\mu}^{\rm vac} = \partial_{\mu}D_{c,\mu\nu} = \partial_{\nu}D_{c,\mu\nu}$.
 
For comparison, note that classical electromagnetic vector potentials can always be written in terms of well defined, classical currents $J_{\mu}$, by an analogous relation : $\displaystyle A_{\mu} = \int\!\!D_{c,\mu\nu}(x-y)\,J_{\nu}$, while the transition to operator QED in the absence of conventional, large scale external fields, involves the replacements of the classical vector potential and currents by operators $A_{\mu}(x)$ and $j_{\mu}(x) = ie\bar\psi(x)\gamma_{\mu}\psi(x)$, which satisfy operator equations of motion : $(-\partial^2) A_{\mu}(x) = j_{\mu}(x)$, or : 
\beq A_{\mu} = \int\!\!D_{c,\mu\nu}(x-y)\,j_{\nu} + \hat A_{\mu}\eeq
where $\hat A$ denotes a free field operator satisfying $(-\partial^2)\hat A = 0$; its vev is zero.

Hence, calculating the vev of $(2.3)$ yields :
\beq <\!0\,|\,A_{\mu}(x)|\,0\!>\, = \int\!\!d^4y\,D_{c,\mu\nu}(x-y)\,<\!0\,|\,j_{\nu}(y)|\,0\!>\eeq
and conventionally, in the absence of the usual, large scale external fields, both sides of $(2.4)$ are to vanish. However, if we assume that non zero $<\!0\,|\,j_{\mu}(x)|\,0\!>$ can exist on ultra short scales, then a comparison of $(2.4)$ with $(2.2)$ suggests that the $A_{\mu}^{\rm vac}(x)$ produced by such small scale currents are to be identified with $<\!0\,|\,A_{\mu}(x)|\,0\!>$ found in conventional QED in the presence of the same $A_{\mu}^{\rm vac}(x)$. In other words :
\beq\matrix{&\displaystyle A_{\mu}^{\rm vac}(x) \,=\,<\!0\,|\,A_{\mu}(x)|\,0\!>\,= \,{1\over i}\,{\delta\over\delta J_{\mu}(x)}\,{\cal Z}[\eta, \bar\eta, J]\Bigr|_{\eta = \bar\eta = J = 0}\hfill\cr\noalign{\medskip} &=\displaystyle {1\over i}\,\int\!\!d^4y\,D_{c,\mu\nu}(x-y){\delta\over\delta A_{\nu}(y)}\,\,e\,^{\displaystyle {\,-{i\over2}\int\!\!{\delta\over\delta A}D_{c}{\delta\over\delta A}}}\hfill\cr\noalign{\medskip} & \displaystyle\times <\!0\,|\,S[A^{\rm vac}]|\,0\!>^{-1}\exp\bigl[ L[A + A^{\rm vac}]\bigr]\Bigr|_{A = 0}\,\hfill}\eeq
which provides a bootstrap equation with which to determine such short scale $A_{\mu}^{\rm vac}(x)$, if any exist. In $(2.5)$, which can be transformed into a functional integral relation, the vacuum to vacuum amplitude is given by {\cite{three}} :
\beq <\!0\,|\,S[A^{\rm vac}]|\,0\!>\,= \,e\,^{\displaystyle {\,-{i\over2}\int\!\!{\delta\over\delta A}D_{c}{\delta\over\delta A}}}\exp\bigl[ L[A + A^{\rm vac}]\bigr]\Bigr|_{A = 0}\eeq
Of course, one immediate solution to $(2.5)$ is $A_{\mu}^{\rm vac}(x) = 0 = <\!0\,|\,A_{\mu}(x)|\,0\!>$, the conventional solution. But we are interested in, and shall find, solutions that may be safely neglected at conventional nuclear and atomic distances, but which are non zero in an interesting way at much smaller distances. 
\bigskip
{\bf\section{Approximation}}
\setcounter{equation}{0}
How does one go about finding a solution to $(2.5)$ ? The first requirement is a representation for $ L[A +  A^{\rm vac}]$ which is sufficiently transparent to allow the functional operation of $(2.5)$ to be performed. Use of the Fradkin functional representation {\cite{five}} for $L[A]$ is one such possibility, because that representation is gaussian in $A$, and $(2.5)$ can be performed immediately. But one is then left with the task of evaluating the remaining functional operations, which is not a trivial affair. What shall be done here is to use the well known, gauge independent, second order perturbative approximation to $L$ :
\beq L[A+A^{\rm vac}] \longrightarrow { i\over2}\int\!\!d^4x\,d^4y\,[A^{\mu}+A^{\mu{\rm vac}}](x)\,K_{\mu\nu}(x-y)\, [A^{\nu}+A^{\nu{\rm vac}}](y)\eeq
which is clearly quadratic in $A$, and corresponds to the simplest Feynman diagram of a virtual lepton--antilepton bubble. The strictly gauge invariant form of that bubble may be represented as :
$$\tilde K_{\mu\nu}(k) = \bigl( \delta_{\mu\nu} \,k^2 - k_{\mu}k_{\nu}\bigr)\,\Pi(k^2)\ ,\ \ \ \ \ k^2 = \vec k^2 - k_0^2$$
and the relevant form of $\Pi(k^2)$ will follow immediately :
\beq\Pi(k^2) = -{2\alpha\over\pi}\!\int_0^1\!\!dy\,y(1-y)\int_0^{\infty}\!\!{ds\over s}\,e\,^{\displaystyle -is[m^2+y(1-y)k^2]}\eeq 
where $m$ denotes the mass of the charged particle whose `` bubble '' is the source of $A^{\rm vac}_{\mu}$. The obvious UV logarithmic divergence of $(3.2)$ will require the usual regularisation and subtraction procedure, and the renormalized $\Pi_R(k^2)$ may easily be put into the form {\cite{six}} :
 \beq\Pi_R(k^2) = {2\alpha\over\pi} \int_0^1\!\!dy\,y(1 - y)\ln\bigl[ 1 + y(1 - y){k^2\over m^2}\bigr]\eeq
Higher order perturbative terms should each yield a less important contribution to the final answer, although the latter could be qualitatively changed by their sum; we shall assume that this is not the case, and that the perturbative approximation ( properly unitarized by the functional calculation which automatically sums over all such loops ) gives a qualitatively reasonable approximation.

Using the approximation $(3.1)$, the functional operation of $(2.5)$, now equivalent to a gaussian functional integration, is immediate \cite{three}. For clarity, we recall in the Appendix the functional formula  that yields the approximate relation for this $A_{\mu}^{\rm vac}(x)$ :
\beq A_{\mu}^{\rm vac}(x) = \int\!\!d^4y\,\Bigl( D_cK\,{1\over 1-D_cK}\Bigr)_{\mu\nu}(x-y)A_{\nu}^{\rm vac}(y)\eeq
or, using its Fourier transform $\tilde A_{\mu}^{\rm vac}(k)$  :
\beq \tilde A_{\mu}^{\rm vac}(k) = \Bigl( \Pi_R(k^2){1\over 1-\Pi_R(k^2)}\Bigr)\,\tilde A_{\mu}^{\rm vac}(k)\eeq
which can be written :
\beq \Bigl(  \displaystyle{1 - 2\Pi_R(k^2)\over 1-\Pi_R(k^2)}\Bigr)\,\tilde A_{\mu}^{\rm vac}(k) = 0\eeq

Non zero solutions to $(3.6)$ may be found in the `` tachyonic '' form \cite{seven} :
\beq \tilde A_{\mu}^{\rm vac}(k) = C_{\mu}(k)\,\delta( k^2 - M^2) = C_{\mu}(k)\,\delta(\vec k^2 - k_0^2 - M^2)\eeq
with $M^2$ such that :
\beq\displaystyle  \Pi_R(M^2) = {1\over2}\eeq
and which serves to determine $M$. Note that a solution of form $C_{\mu}\,\delta(k^2 + \mu^2)$ would not be possible, since the $\log$ of $\Pi_R(-\mu^2)$ picks up an imaginary contribution for time--like $k^2 = -\mu^2$, for $\mu> 2m$. $\Pi_R(M^2)$ must be a real quantity to satisfy $(3.8)$.
\bigskip
{\bf\section{Computation}}
\setcounter{equation}{0}
In order to fully describe this $A_{\mu}^{\rm vac}$, one must define $C_{\mu}(k)$. Remembering that the simplifying choice of a Lorentz gauge has already been made for this vacuum potential, one might expect  to be able to choose : 
$$C_{\mu}(k) = v_{\mu} - k_{\mu}\bigl(k_{\nu}v_{\nu}\bigr)/k^2$$
where $v_{\mu}$ is a constant to be determined. But the part proportional to $k_{\mu}$ is a pure gauge term, which cannot contribute to any electromagnetic field, and is therefore irrelevant, while the replacement of $C_{\mu}(k)$ by $v_{\mu}$ serves to generate fields that diverge in the region of the light cone. This is the form of solution used in ref.\cite{one}, which required an ad hoc cut--off when computing energies.

A far better solution is obtained by enforcing the Lorentz gauge condition with the replacement of $C_{\mu}(k)$ by $\kappa v_{\mu}\,\delta(k\cdot v)$ where $\kappa$ is a constant to be determined; this choice produces a vacuum field that is both simple and everywhere finite. But before such a solution can be taken seriously, there are certain questions that must be answered :

a) What is $v_{\mu}$ ? Physically, that vector should represent an electric field polarization, corresponding to the fluctuating electric field in the plane of the fluctuating bubbles. But the QED vacuum will have bubbles fluctuating in all possible planes, or even on curved surfaces; and there can be no $\vec v$ direction singled out. It is then intuitively clear that $v_{\mu}$ should have only a fourth component.

b) But in which Lorentz frame ? If it is to represent a field generated by the same vacuum processes in every frame, it should have the same value to each observer in his own Lorentz frame.

We now show that this choice of solution for the vacuum field does satisfy both of these requirements, and has just the correct behavior to suggest a mechanism for Inflation, while producing a present day energy density that can be associated with Dark Energy.

Insert a representation for both delta functions of :
\beq \tilde A_{\mu}^{\rm vac}(k) =\kappa\, v_{\mu}\,\delta(k\cdot v)\,\delta(k^2 - M^2) \eeq
and calculate the inverse Fourier transform of $\tilde A_{\mu}^{\rm vac}$. It will be represented by the parametric, ``~proper time '' integral :
\beq A_{\mu}^{\rm vac}(x) = {\kappa\over(2\pi)^4}\,\sqrt{i\pi\over4}\Bigl({-i v_{\mu}\over \sqrt{v^2}}\Bigr)\int_{-\infty}^{+\infty}\!\!ds\,s^{-3/2}\,\epsilon(s)\,e\,^{\displaystyle isM^2 + iX^2/4s}\eeq
with $\displaystyle\epsilon(s) = {s\over|s|}$, $x\cdot v = \vec r\cdot\vec v - x_0v_0$, and where $X^2 = x^2 -(x\cdot v)^2/v^2$, and is a Lorentz invariant quantity. The corresponding solution in another Lorentz frame, represented by a prime on $x$ and a prime on $v$, will have exactly the same form.

Now, consider the solution $(4.2)$ in our frame, and ask what value should be assigned to the spatial components. From the argument of a) above, the only sensible value for this field $A_{\mu}^{\rm vac}$ is the choice $\vec v = 0$. Then, the quantity $X^2$ reduces to $r^2$, and $(4.2)$ can be evaluated trivially :
\beq A_{\mu}^{\rm vac}(x)\longrightarrow A_4^{\rm vac}(x) = i\,{\kappa\over(2\pi)^3}\,\epsilon(v_0)\,{\sin(Mr)\over r}\eeq
and depends only on $r$ and on the sign of $v_0$.

Now switch to another Lorentz frame, where the result is given by $(4.2)$ using prime variables, related to the unprimed variables by standard Lorentz transformation. An observer in that frame asks what value he should assign to the spatial components of his $v'$; and for his vacuum field, he comes to exactly the same conclusion as did we : the only sensible choice for a vacuum field as seen by him must require $\vec v\,' = 0$. Note that his square root variable $\displaystyle\sqrt{X^2}$ containing all the $x'$ dependence is a Lorentz invariant quantity, and is equal to our square root variable. In our frame, when we set $\vec v = 0$, that variable reduces to $r$; and in his frame, when he sets $\vec v\,' = 0$ it reduces to $r'$. But both must be equal, since they were derived from the same invariant; when the observer in the primed frame sets his $\vec v\,' = 0$, as he must to describe his vacuum field, he is using the same functional form as $(4.3)$ in terms of his $r'$.

The only possible difference between the two expressions of $(4.3)$, primed or unprimed, is the sign of $v_0$ and that of $v'_0$. But, as always when dealing with physical entities, we restrict all admissible Lorentz transformations to those which are orthochronous, keeping the same sense of time, or of energy, or in this case of $v_0$; and hence $\epsilon(v_0) = \epsilon(v'_0)$ and the two versions of $(4.3)$ are the same. In this way, observers in every Lorentz frame see the same vacuum field. For simplicity, we shall choose $\epsilon(v_0) = +1$, although this choice of sign has no bearing on the vacuum energy densities to be calculated.

Following the arguments above, any observer in any frame will see a `` vacuum electrostatic '' potential of form  :
$$\displaystyle\phi^{\rm vac}(r) = {\kappa\over(2\pi)^3}\,{\sin(Mr)\over r}$$
where $\kappa$ is a constant to be determined. 

The resulting electric field has a rapid spatial variation, as does its energy density :
\beq\rho = \vec{\cal E}\,^2/8\pi = \xi\,{M^2\over r^2}\Bigl( \cos(Mr) - {\sin(Mr)\over Mr} \Bigr)^2 = \xi\,M^4f(x)\eeq
where $\displaystyle\xi = ({1\over8\pi}){\kappa^2\over(2\pi)^6}$, $x = Mr$, and $\displaystyle f(x) = {1\over x^2}\Bigl( \cos x - {\sin x\over x} \Bigr)^2$. A plot of $f(x)$ has the form indicated in Fig.1, which suggests the basic idea of this approach : the first pulse serves to kick start Inflation, which is supposed to begin at $t\sim 10^{-42}$ s, and have an average energy density $\rho$ such that $\rho^{1/4}\sim 10^{18}$ GeV. These numbers, and the time when Inflation stops, $t\sim 10^{-32\pm 6}$ s, with an average $\rho^{1/4}\sim 10^{13\pm 3}$ GeV, are from Table $2.1$ of Liddle and Lyth \cite{eight}. However, the initial $\rho^{1/4}$ has simply been specified as the Planck mass, with no uncertainties attached, for the Planck mass just symbolizes the beginning of Inflation; in reality, several orders of magnitude of uncertainties should be associated with that number, or with any such number for which a model exists. Again, the relation of our model to Inflation -- in particular, to the first peak of Fig.1 -- will be discussed  elsewhere,  in detail.

\begin{figure}
\includegraphics[width=10truecm]{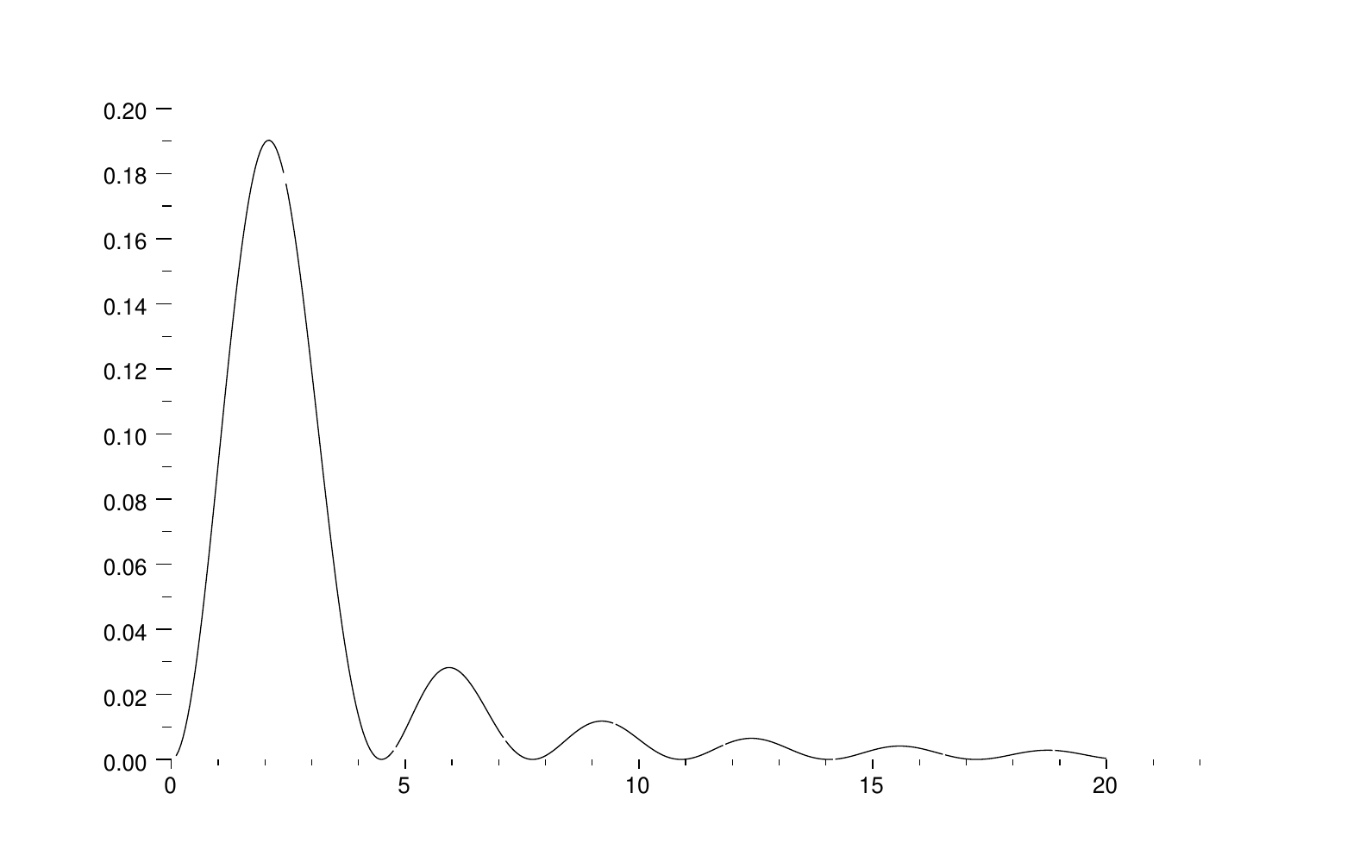}
\caption{A plot of $f(x) = \displaystyle{1\over x^2}\Bigl(\cos x - {\sin x\over x}\Bigr)^2$ vs. $x$}
\end{figure}

\bigskip
{\bf\section{Application to Dark Energy}}
\setcounter{equation}{0}
The first question to be decided is which are the fundamental, quantized fields of Physics whose quanta are electrically charged, and are appropriate to be included. The leptons and the quarks are the natural, fundamental fermions out of which the charged particles forming the Standard Model are built; and it is these particles whose simplest closed loops will form the $A_{\mu}^{\rm vac}$  relevant to present day Dark Energy.

We simplify the analysis by neglecting the $1$ within the logarithms of (3.3), and replace the average of the three lepton masses and of the six quark masses by 1 GeV. These approximations are for convenience and simplicity only, and have no real bearing on the results. The relevant equation determining $M$ is then : 
 \beq\displaystyle{1\over2} = \Pi_R(M^2) = {2\alpha\over3\pi} \biggl[ 3\ln\Bigl({M\over \bar m_l}\Bigr) + 3\times3\times({4\over9} + {1\over9})\ln\Bigl({M\over \bar m_q}\Bigr)\biggr]\eeq
With $\bar m_l = \bar m_q = m_0 = 1\,$GeV, $\alpha = 1/137$ and a color factor of 3 being included, this equation reduces to :
 \beq\displaystyle{1\over2} = \Pi_R(M^2) = {2\alpha\over3\pi} \biggl[ 8\ln\Bigl({M\over m_0}\Bigr) \biggr]\eeq
and we find  :
\beq M = 10^{18}\,{\rm GeV}/c^2\eeq
It is interesting to note that this perfectly finite calculation, in the context of QED, is able to produce a mass term on the order of the Planck mass. 

With this value of $M$, we are then able to compute the order of magnitude of present day vacuum energy. If  we compute that energy density by integrating the energy density of this vacuum field, as expressed in $(4.4)$ -- or more simply, by extracting the dominant, non oscillatory behavior of that integral -- and dividing by the present day volume of the Universe, $(4\pi/3)R_{\rm now}^3$, one finds :
\beq \rho_{\rm now}\sim {3\over2}\,\xi\,{ M^2\over R_{\rm now}^2} \eeq
In order to obtain the density in ${\rm g/cm}^3$, we choose to restore the dimension  using the couple of parameters relevant to this mechanism : the speed of light and the electric charge. It gives : 
\beq \rho_{\rm now}\sim \,\xi\,{c^2\over e^2}\,{ M^2\over R_{\rm now}^2} \eeq
Using :
$M = 10^{18}\,$GeV/$c^2$ $ =2.\ 10^{-9}\,$kg, $R = 4.\ 10^{26}\,$m, $c = 3.\ 10^8\,$ms$^{-1}$ and  $e^2 = 2.\ 10^{-28}\,$Nm$^{2}$, the result is :
\beq \rho_{\rm now}\sim \xi\,10^{-29}\,{\rm g/cm}^3\eeq
which, choosing $\xi\sim O(1)$ is precisely the order of magnitude needed to fit the acceleration data.

This model of a QED vacuum energy, defined by a particular choice of $C_{\mu}(k)$, is, of course, not the only possibility; but it is everywhere finite, with a simple Lorentz invariance form. 
\bigskip
{\bf\section{Summary}}
\setcounter{equation}{0}
The above Sections have defined a simple, intuitive account for Dark Energy based upon a new, boostrap model of a QED vacuum energy. Use of the simplest lepton loop fluctuations allowed a Gaussian functional operation to be performed; but there are surely small, perturbative corrections to our  equation for $A_{\mu}^{\rm vac}$. Quite apart from numerical approximations, the basic idea of such a QED vacuum field and source of residual energy, well defined and everywhere finite, represents a slight enlargement of standard QED, one which cannot be negated by any test at current particle energies. It is a new possibility.

\vskip1truecm {\bf\Large Acknowledgments} 
\vskip0.3truecm 
It is a pleasure to acknowledge helpful conversations with Walter Becker, Ian Dell'Antonio, and Savvas Koushiappas.

This publication was made possible through the support of a grant from the John Templeton Foundation. The opinions expressed in this publication are those of the authors and do not necessarily reflect the views of the John Templeton Foundation.
\bigskip
\bigskip
{\bf\section* {Appendix}}
\setcounter{equation}{0}
\appendix
Defining the linkage operator as :
$$\exp{\cal D} = \exp\bigl[-{i\over2}\int\!\!dx\,dy\,{\delta\over\delta j(x)}A(x,y){\delta\over\delta j(y)}\bigr]$$
we have the following relation \cite{three} :
$$\displaylines{\exp{\cal D}\exp\bigl[{i\over2}\int\!\!du\,dv\,j(u)B(u,v)j(v) + i\!\int\!\!du\,f(u)j(u)\bigr]\cr  = \exp\bigl[{i\over2}\int\!\!du\,dv\,dx\,j(u)B(u,x)( 1 - AB )^{-1}(x,v)j(v) + i\!\int\!\!du\,dv\,f(u)( 1 - AB )^{-1}(u,v)j(v)\bigr]\cr +{i\over2}\int\!\!du\,dv\,dx\,f(u)A(u,x)( 1 - BA )^{-1}(x,v)f(v)-{1\over2}\,{\rm Tr}\log( 1 - AB )\bigr]\cr}$$
This formula, used in eq.(2.5) with the approximation (3.1)  leads to eq.(3.4).

\end{document}